\documentclass[10pt,amstex]{article}

\usepackage[usenames]{color}
\usepackage{amscd}
\usepackage{amsmath}
\usepackage{amssymb}
\usepackage{graphicx}

\newtheorem {theo} {\bf Theorem} [section]
\newtheorem {prop} [theo] {\bf Proposition}
\newtheorem {cory} [theo] {\bf Corollary}
\newtheorem {lem} [theo] {\bf Lemma}
\newtheorem {defn} [theo] {\bf Definition}

\newtheorem {rem} [theo] {\bf Remark}
\newcommand{\QED}{\hfill \CaixaPreta \vspace{6mm}}
\def\CaixaPreta{\vrule Depth0pt height6pt width6pt}

\newcommand{\qed}{\nopagebreak\hfill{\vrule width6pt height6pt depth0pt}}
\newcommand{\dpy}{\displaystyle}
\newcommand{\be}{\begin{eqnarray}}
\newcommand{\ee}{\end{eqnarray}}
\newcommand{\benn}{\begin{eqnarray*}}
\newcommand{\eenn}{\end{eqnarray*}}
\newcommand{\bse}{\begin{equation}}
\newcommand{\ese}{\end{equation}}
\newcommand{\bsenn}{\begin{displaymath}}
\newcommand{\esenn}{\end{displaymath}}
\newcommand{\ot}{\otimes}
\newcommand{\spec}{\;{\rm spec }\;}
\newcommand{\logand}{\;\;{\rm and }\;\;}

\newcommand{\logor}{\;\;{\rm or }\;\;}

\newcommand{\where}{\;\;{\rm where }\;\;}

\newcommand{\R}{\mbox{I${\!}$R}}

\begin{document}

\title{Stability of Large Flocks: an Example}
\author{ J. J. P. Veerman\thanks{e-mail: veerman@pdx.edu} (1,2)\\
%, F. M. Tangerman (3),\\
(1) Dept. of Math. \& Stat., Portland State
University, Portland, OR 97201, USA. \\
(2) Center for Physics and Biology, Rockefeller University, NY, NY 10021.\\
%(3) Northport, NY.
}
\maketitle

\begin{abstract}
The movement of a flock with a single leader (and a directed path from it to every agent) can be stabilized.
Nonetheless for large flocks perturbations in the movement of the leader may grow to a considerable size as they propagate throughout
the flock and before they die out over time.
As an example we consider a string of $N+1$ oscillators moving in $\R$. Each one `observes' the relative velocity and position of only its nearest neighbors. This information is then used to determine its own acceleration. Now we fix all parameters \emph{except} the number of oscillators. We then show (within a certain class of systems) that
a perturbation in the leader's orbit is almost always amplified \emph{exponentially} in $N$ as it propagates towards the outlying members of the flock. The \emph{only} exception is when there is a symmetry present in the interaction: in that case the growth of the perturbation is \emph{linear} in $N$.
\end{abstract}

 %\normalsize        %\mysetfontsize5

\begin{centering}\section{Introduction} \label{chap:intro} \end{centering}
\setcounter{figure}{0} \setcounter{equation}{0}

\begin{figure}[ptbh]
\centering
\includegraphics[width=4.6in]{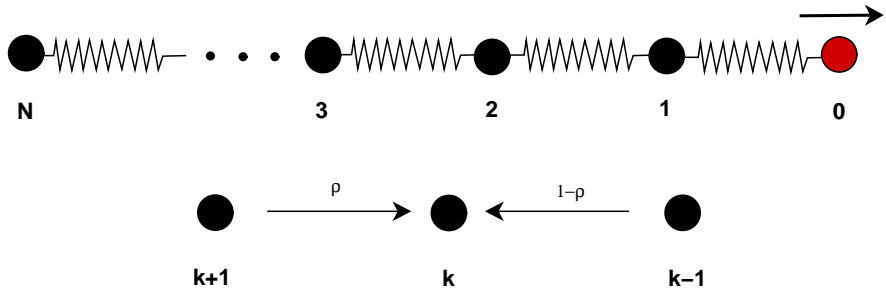}
\caption{\emph{ In the upper figure the `symmetric' ($\rho=\frac12$) flock is illustrated.
Each agent is linearly coupled to its nearest neighbor. At $t=0$ the agent labeled 0, the leader, undergoes a forced motion: either an oscillation or a kick in the direction of the arrow above it. It receives no feedback from the flock.
The asymmetric interaction is suggested in the lower figure, where we drew the $k$-agent with its interactions and their weights. The arrows give the direction of the information flow. }}
\label{fig:chain}
\end{figure}

We take up the study of a system very familiar from physics (\cite{seitz} for example): a long string of coupled linear oscillators (called `agents' or `cars' from here on) with linear nearest neighbor interactions. In order to model things like traffic we change certain features of the model that is so familiar in the context of statistical physics. The most significant ones are that we do not assume that interaction is symmetric
and that the oscillators are damped. Furthermore we allow the system to have non-zero mean velocity. Finally we assume that the chain is \emph{finite} and we take into account boundary effects from  the ends of the chain (no periodic boundary conditions). The class of systems whose study we take up in this paper is illustrated in Figure \ref{fig:chain} and made precise in Equation \ref{example2}.

The purpose here is to gain a qualitative understanding of how a perturbation in the movement of one the agents (which we call the \emph{leader}) propagates throughout the flock of agents.
Here is a more precise formulation. Suppose that a flock is moving coherently or ``in formation", ie: each member has the same constant velocity, and inter-agent distances are constant. Now the leader's motion is perturbed. After a while also the agent farthest away from the leader (called the \emph{trailing car}) senses the effect, and its orbit will start to deviate from the in formation orbit. What is the ratio between these two perturbations \emph{as a function of the size $N$ of the flock}, while keeping all other parameters fixed ?

The immediate motivation is to study which kinds of interactions might take place in actual large flocks or might seem desirable to implement in artificial flocks (think of cars on a highway equipped with an automatic pilot). It would seem that in both types of systems the kinds of interaction preferred would be those where the growth rate just mentioned is as low as possible.

Ideally we find the growth rate of perturbations
in many different models, for example where interactions are allowed to include up to 2 or 3 or 10 neighbors on either side. This seems analytically out of reach. Thus the acceleration of the $k$-th car is determined by its observations of the $k-1$-st and the $k+1$-st car. What we chose to do instead is to introduce a \emph{weighted} nearest neighbor interaction, that is: For $\rho\in[0,1]$ we weight the information coming from the $k-1$-st car with $1-\rho$, and the other with  $\rho$ (see Equation (\ref{example2})).

This line of thought has many potential applications in technology. Suppose for example one wants to automatically control dense traffic on a single lane road. Each car is equipped with sensors that register the relative velocity and position of its neighbors. This information is then used to regulate the acceleration of the cars.
A typical example is the `canonical traffic problem' proposed in \cite{flocks4} and \cite{flocks5}, where the lead car abruptly accelerates (when, for example, a traffic light turns green), and the other cars try to follow.

Our conclusions for the systems typified by Equation \ref{example2}, can be summarized as follows. In all systems the ratio of the size of perturbation of the trailing car to the perturbation of the leader grows exponentially in the size $N$ of the flock, with one exception: when $\rho=\frac12$. In the last case (called the \emph{symmetric} case because equal attention is paid to the cars on either side) this growth is only \emph{linear} as function of $N$. The study of the symmetric case was reported in \cite{flocks4, flocks5}. Here we continue that program by investigating the asymmetric case.

The curious fact that perturbations in the leader's position or velocity is necessarily unbounded
as the size of the flock grows is of course
of paramount importance in many applications (automated traffic, biological flocks, and so on).
In fact this has been commented upon by several authors already, notably \cite{SPH,BH} in cases
similar to our $\rho=0$ and $\rho=\frac12$ cases. In this note we give precise estimates for
how these perturbations grow as function of the number of agents,
not only in those cases but also for other models (all $\rho\in[0,1]$).

A more distant motivation to study these systems is to attain a general understanding of linear oscillators interacting according to a more general graph (called the ``communication graph", see for instance \cite{flocks2} for definitions). In that context similar questions arise but in a more general context. What we present here is a simple example.

Section \ref{chap:model} defines the model. In Section \ref{chap:stability} the definitions of stability we use are given and briefly discussed. In it we also state the main result of this note. Sections \ref{chap:asymptotic} and \ref{chap:harmonic} prove these results. (Some of the more calculational steps in these proofs are relegated to the Appendix.)

\vskip .2in
\noindent{\bf Notational Conventions:} To avoid confusion, we finally list two important conventions here. The first is that after Section \ref{chap:asymptotic} we assume that both $f$ and $g$ are negative reals to insure asymptotic stability (Theorem \ref{theo:stable}). The second is that in order for certain expressions ($\mu_+(i\omega)$ and $\mu_-(i\omega)$, for $\omega>0$) in Theorem \ref{theo:a_n}) to be continuous functions it is convenient to define the symbol $\sqrt z$ as the root with angle in the interval $[0,\pi)$ (branch cut along the positive real axis).

\vskip .2in
\section*{Acknowledgements:} I am grateful for useful conversations with Folkert Tangerman.

\vskip 1.in
\begin{centering}\section{The Model} \label{chap:model} \end{centering}
\setcounter{figure}{0} \setcounter{equation}{0}

We define the model and propose a notion of stability for flocks. Our strategy is to study qualitative aspects of the solution for fixed $\rho$, $f$, and $g$, and as we let $N$ tend to infinity.

Let $f$ and $g$ be real, and $\rho\in[0,1]$.
The model is given by:
 \be
\forall i \in \{1,\cdots N-1\} \;:\; \dot x_i &=& u_i \cr
\dot u_i &=&f\left\{(x_{i}-h_{{i}})- (1-\rho)(x_{i-1}-h_{{i-1}})-\rho(x_{i+1}-h_{{i+1}})\right\}
+g\left\{u_i-(1-\rho)u_{i-1}-\rho u_{i+1}\right\}\cr
 \dot x_{N} &=& u_{N} \cr \dot u_{N}
&=& f\left\{(x_{N}-h_{{N}})- (x_{N-1}-h_{{N-1}})\right\}+
g\left\{u_{N}-u_{N-1}\right\}\cr
 \logand \quad \quad  x_0 &=& x_0(t) \quad {\rm given}
 \label{example2}
 \ee
The (constant) parameters $h_i$ determine the desired relative distances between agents $i$ and $i-1$ as $h_i-h_{i-1}$. The feedback parameters $f$ and $g$ are independent of $i$ and time. Note that the total number of agents is in fact also a parameter in this problem. We do not carry this into the notation. The subscript $N$ will always stand for the last agent in a system with agents numbered from 0 to $N$. Similarly we do not carry the dependence on the parameters $f$ and $g$ explicitly into the notation.

One can show that the orbits of this system such that the distances between successive cars are preserved (namely: $-(h_i-h_{i-1})$) form a 2-parameter family, namely $x_k(t)=x_0(0)+v_0(0)t+h_k$ and $\dot x_k(t)= v_0(0)$. We will call these orbits \emph{in formation orbits}.

It is advantageous to write Equation (\ref{example2}) in a more compact form. Introduce the notation
\bsenn
z\equiv (z_1,\dot z_1, z_2,\dot z_2,\cdots, z_N,\dot z_N) \equiv
(x_1-h_1,u_1,x_2-h_2,u_2, \cdots, x_{N}-h_N, u_{N})^T \quad .
\esenn
The leading car is not encoded since its orbit is a priori given. The system can now be recast as a first order ODE:
\bse
\dot z = M z + \Gamma_0(t)\quad .
\label{eq:indepleader}
\ese
The details of this are discussed in \cite{flocks2}, here we just give the relevant definitions.

Let $I$ and $P$ are $N$-dimensional square matrices, where $I$ is the identity and $P$ is given by
\bse
P= I-Q_\rho \quad \where\quad Q_\rho=\left(\begin{array}{ccccc}
0 & \rho & & & \\
1-\rho & 0 & \rho & & \\
 & \ddots & \ddots &  \ddots & \\
 & & 1-\rho & 0 & \rho \\
 & & & 1 & 0
\end{array}\right),
\label{eq:laplacian}
\ese
$P\equiv I-Q_\rho$ is called the reduced graph Laplacian. It describes  the flow of information among the agents, with the exception of the leader (hence the word `reduced'). The $2\times2$ matrices $A$ and $K$ are given by:
\bse
A= \left(\begin{array}{cc}
0 & 1 \\
0 & 0
\end{array}\right) \quad \logand \quad
K= \left(\begin{array}{cc}
0 & 0 \\
f & g
\end{array}\right) \quad .
\label{eq:A-and-K}
\ese
The orbit of the leader is assumed to be a priori given and therefore only appears in the forcing term $\Gamma_0(t)$.
We will refer to this agent as an \emph{(independent) leader}.
Analyzing Equation (\ref{example2}) and assuming without loss of generality that $h_0=0$, one gathers that:
\bse
\Gamma_0(t) = \left(\begin{array}{c}
0  \\
(1-\rho)\left(fz_0(t)+g\dot z_0(t)\right) \\
0\\
\vdots
\end{array}\right) \quad  .
\label{eq:Gamma_0}
\ese
To define $M$ of Equation (\ref{eq:indepleader}) in terms of these quantities, we use the Kronecker product ($\otimes$)
 \bse
M\equiv I \ot A  + P \ot K \quad  .
 \ese
The advantage of this somewhat roundabout way of defining the matrix $M$ is that in the eigenvalues of the reduced Laplacian $P$ in many cases are known. From that the eigenvalues of $M$ can then be derived.
That is the program followed in the Section \ref{chap:asymptotic}.

\vskip 1.in
\begin{centering}\section{Stability of Flocks} \label{chap:stability} \end{centering}
\setcounter{figure}{0} \setcounter{equation}{0}

\noindent
\begin{defn}
The system given in Equation \ref{eq:indepleader} is called `asymptotically stable' if all eigenvalues of $M$ have negative real part.
\label{defn:asymptotical}
\end{defn}
Suppose for the moment that $\Gamma_0(t)=0$ fot $t>t_0$. Then
the solution of the system tends to 0 exponentially fast (in $t$) if and only if the system is asymptotically stable. This corresponds to the usual notion of asymptotic stability (see for example \cite{arnold}, Section 23). The question when the system is asymptotically stable has a straightforward answer (see Theorem \ref{theo:stable}): it is if and only if $f$ and $g$ in Equation (\ref{example2}) are negative.
We will therefore from now on \emph{assume that $f$ and $g$ are negative}.

One can show (Proposition \ref{prop:a_k}) that if the leader executes an oscillation of the form $e^{i\omega t}$, then $z_k(t)$ tends to $a_k(i\omega)e^{i\omega t}$ as $t$ tends to infinity. The functions  $a_k(i\omega)$ are called the frequency response functions.

\begin{defn} Let $A_N\equiv \sup_{\omega\in\R}\; |a_N(i\omega)|$. The system is called `harmonically stable' if it is asymptotically stable and if $\limsup_{N\rightarrow \infty}\; \left|A_N\right|^{1/N}\leq 1$. Otherwise the system is called `harmonically unstable'.
\label{defn:harmonic}
\end{defn}
This kind of instability roughly says that certain long-term oscillatory perturbations in the orbit of the leader will have their amplitude magnified by a factor that is exponentially large in $N$.
In Theorem \ref{theo:stable2} we establish that the system is harmonically unstable if $\rho\neq 1/2$; harmonic stability for $\rho=\frac12$ was established in \cite{flocks4}.

In earlier work (\cite{flocks5}) we studied a `fundamental traffic problem' which roughly corresponds to setting the acceleration of the leader equal to the Dirac delta function, $\delta(t)$, and $ z_0(0)=\dot z_0(0)=0$. This gives $z_0(t)=t$ for $t\geq 0$ which can be substituted into Equation (\ref{eq:Gamma_0}).

\begin{defn} Consider Equation (\ref{eq:indepleader}) with forcing determined by $\ddot z_0(t)\equiv \delta(t)$ and subject to the initial conditions $z_k(0)=\dot z_k(0)=0$ and $\dot z_k(0)=0$. Let $Z^{(i)}_N\equiv \sup_{t>0} |\frac{d^i}{dt^i}(z_N(t)-z_0(t))|$. The system is called `impulse stable' if it is asymptotically stable and if for $i$ 0, 1 and 2, we have $\limsup_{N\rightarrow \infty}\; \left|Z^{(i)}_N\right|^{1/N}\leq 1$. (And `impulse unstable' in the other case.)
\label{defn:impulse}
\end{defn}
Loosely interpreted this kind of instability means that if we give the leader a 'unit-kick', then that perturbation travels through the flock and causes $\sup|x_N(t)|$, $\sup |\dot x_N(t)|$, or $\sup |\ddot x_N(t)|$ to grow exponentially in $N$, before eventually dying out (due to asymptotic stability).

Impulse stability is perhaps at first sight more natural or appealing than harmonic stability because it is formulated in the time domain, whereas the latter takes place in the Fourier domain. However mathematically the criterion is much harder to check.
In \cite{flocks5} we proved that the case $\rho=1/2$ is impulse stable. But the case $\rho\neq 1/2$ is much more problematic and will be taken up in a separate work (\cite{flocks7}).

It therefore may be argued that all three kinds of stability are necessary to form large flocks. From the above remarks, it follows that of the systems investigated here only the one with $\rho=\frac12$ satisfies all three.
It is interesting that even in that case \emph{linear} growth of $Z^{(i)}_N$ still takes place (see \cite{flocks4, flocks5}). This seems to be the best case possible. we summarize this discussion with our main result.

\begin{theo} The system given by Equation (\ref{example2}) satisfies all three stability criteria if and only if $f$ and $g$ are negative and $\rho=1/2$.
\label{theo:main}
\end{theo}

\vskip 1.in
\begin{centering}\section{Asymptotic Stability} \label{chap:asymptotic} \end{centering}
\setcounter{figure}{0} \setcounter{equation}{0}

\noindent We derive the criterion for asymptotical stability.

In the statement of the next result we use the following equation, where $\rho \in(0,1)$ and $\phi$ are real variables:
\begin{equation}\label{cot2}
(2\rho - 1)\cot\phi=\cot N\phi\; .
\end{equation}
Recall that the matrix $P$ is defined in Equation (\ref{eq:laplacian}).

\vskip.2in\begin{prop} (\cite{tridiagonal})
For any $\rho\in(0,1)$, the matrix $P$ has $N$ distinct eigenvalues $\{\lambda_\ell\}_{\ell=0}^{N-1}$: \\
\emph{\bf i) If $\rho\in (0,\frac{1}{2}]$:} for $\ell\in\{0,\ldots, N-1\}$, $\lambda_\ell=1-2\sqrt{\rho(1-\rho)}\,\cos \phi_\ell$, where $\phi_\ell \in \left( \frac{\ell\pi}N, \frac{(\ell+1)\pi}N\right)$ solves (\ref{cot2}).\\
\emph{\bf ii) If $\rho\in (\frac12,\frac{N+1}{2N}]$:} Identical to i).\\
\emph{\bf iii) If $\rho\in (\frac{N+1}{2N},1)$:} for $\ell\in\{1,\ldots, N-2\}$, $\lambda_\ell=1-2\sqrt{\rho(1-\rho)}\,\cos \phi_\ell$, where $\phi_\ell \in \left( \frac{\ell\pi}N, \frac{(\ell+1)\pi}N\right)$ solves (\ref{cot2}); $\lambda_0=\frac{(2\rho-1)^2}{2\rho^2}\left(\frac{1-\rho}{\rho} \right)^{N-1} + \mathcal{O}\left(\left(\frac{1-\rho}{\rho} \right)^{2N-2}\right)$ and $\lambda_{N-1}=2-\lambda_0$.
\label{prop:evalsP}
\end{prop}

\noindent One can show (see \cite{flocks2,flocks4, flocks5}) that the eigenvalues of $M$ defined in Equation (\ref{eq:indepleader}) are given by the solutions $\nu_{\ell\pm}$ of
 \bse
 \nu^2-\lambda_\ell g \nu-\lambda_\ell f= 0  \quad,
  \label{eq:evals2}
 \ese
where $\lambda_\ell$ runs through the spectrum of $P$. So:

\begin{theo} The eigenvalues of $M$ are
\bsenn
\nu_{\ell\pm} = \frac{1}{2}\left(\lambda_\ell g \pm \sqrt{(\lambda_\ell g)^2 + 4 \lambda_\ell f}\right)= \frac{\lambda_\ell g}{2}\left(1\pm \sqrt{1+\frac{4f}{\lambda_\ell g^2}}\right) \quad ,
 \esenn
where $\lambda_\ell$ runs through the spectrum of $P$. Because the $\lambda_\ell$ are contained in the interval $[0,2]$ (see Proposition \ref{prop:evalsP}), the system is stabilized (or asymptotically stable) if and only if both $f$ and $g$ are strictly smaller than zero.
\label{theo:stable}
\end{theo}

\vskip 1.in
\begin{centering}\section{Harmonic Stability} \label{chap:harmonic} \end{centering}\
\setcounter{figure}{0} \setcounter{equation}{0}

In this section we calculate (Proposition \ref{theo:a_n}) and study the properties (Theorem \ref{theo:stable2}) of the frequency response function of the trailing car in detail. The main result of this Section is  Theorem \ref{theo:stable2} that says that if $[0,1)\backslash\{\frac12\}$ then the system is harmonically unstable. (If $\rho=1$ the question is moot as the leader's motion goes unperceived by the flock.)

The following constant will frequently simplify formulae:
\bsenn
\kappa \equiv \frac{1-\rho}{\rho} \quad \logor \quad \rho = \frac{1}{1+\kappa} \quad .
\esenn

\vskip .2in
\begin{prop} If $z_0(t) = e^{\nu t}$ and $\nu\in i\R$ and $\nu \not\in \spec(M)$, there is a sequence $a=(a_1\cdots, a_N)$ of complex numbers $\{a_k(\nu)\}_{k=1}^N$ so that trajectory of the system is asymptotic to:
\bsenn
z_k(t) = a_k(\nu)\,e^{\nu t} \quad \where \quad a(\nu)= -(M-\nu I)^{-1}g_0.
\esenn
\label{prop:a_k}
\end{prop}

\noindent {\bf Proof:} (See also \cite{flocks2}.) Since $z_0(t) = e^{\nu t}$, the non-autonomous term in Equation (\ref{eq:indepleader}) can be written as $\Gamma_0(t)=g_0e^{\nu t}$, where $g_0$ is a constant vector. The general solution of the system is: $e^{Mt}z_0+(M-\nu I)^{-1}\left( e^{Mt}-e^{\nu t}\right)g_0$, and $e^{Mt}$ tends to zero. \QED

\noindent The complex functions $a_k(\nu)$ (where $\nu \in i\R$) are called the \emph{frequency response} (of the $k$-th agent).

\vskip .2in
\begin{prop}
i): For $\rho\in (0,1)\backslash \{\frac{1}{2}\}$ the frequency response function of the $k$-th agent is given by the functions:
 \bsenn
\begin{array}{c}
\dpy a_{k}(\nu)=
\kappa^k  \frac{\left(\mu_+ -\mu_+^{-1}\right)\mu_+^{N-k}- \left(\mu_- - \mu_-^{-1}\right)\mu_-^{N-k}}
{\left(\mu_+ -\mu_+^{-1}\right)\mu_+^{N}- \left(\mu_- - \mu_-^{-1}\right)\mu_-^{N}} \\
\where \quad \dpy \mu_\pm=\mu_\pm(\nu) \equiv \frac{1}{2\rho}\left(\gamma\pm \sqrt{\gamma^2-4\rho(1-\rho)}\right) \quad \logand \quad
\gamma = \gamma(\nu)\equiv \dpy \frac{f+ g \nu -\nu^2}{f+ g\nu}
\quad .
\end{array}
 \esenn
 ii): (\cite{flocks5}) When $\rho=\frac12$ the above expressions simplify to:
 \bsenn
\begin{array}{c}
\dpy a_{k}(\nu)=\frac{\mu_+^{N-k}+\mu_-^{N-k}} {\mu_+^{N}+\mu_-^{N}} \\
\where \quad \mu_\pm(\nu) \equiv \gamma\pm \sqrt{\gamma^2-1} \quad \logand \quad
\gamma = \gamma(\nu) = \dpy \frac{f+ g \nu - \nu^2}{f+ g \nu}
\quad .
\end{array}
\esenn
iii): (\cite{flocks4}) When $\rho=0$:
\bsenn
a_k(\nu)= \gamma(\nu)^{-k} \quad .
\esenn
 \label{theo:a_n}
\end{prop}

\noindent {\bf Proof:} The reasoning of i) is identical to that in Lemma 3.2 of \cite{flocks5}. \QED

\noindent{\bf Remark:} Even though $\mu_\pm$ are not rational functions of $\nu$, one can check that in fact the $a_k(\nu)$ in fact are proper rational.

\vskip .1in
\noindent{\bf Remark:} The coefficients $\mu_\pm$ in the Theorem are the roots of the following quadratic equation:
\bsenn
\rho \mu^2-\gamma \rho +(1-\rho) = 0 \quad .
\esenn

\vskip .2in
The most important case of Theorem \ref{theo:a_n} is the frequency response of the trailing car, labeled $N$, when $\rho\in (0,1)\backslash \{\frac12 \}$. The previous result immediately implies:

\begin{cory} For $\rho\in (0,1)\backslash \{\frac{1}{2}\}$ the frequency response function of the last agent is given by
\bsenn
a_{N}(\nu)=  \frac{1+\kappa}{\kappa} \;\kappa^N \; \frac{\mu_+ - \mu_-}
{\left(\mu_+-\mu_+^{-1}\right)\mu_+^{N}- \left(\mu_--\mu_-^{-1}\right)\mu_-^{N}} \quad ,
\esenn
with $\mu_\pm$ and $\gamma$ as before.
\label{cor:trailing}
\end{cory}

Because the inverse Fourier Transform of $a_N$ equals the real valued function $\ddot z_N(t)$, we are mainly interested in the case where $\nu=i\omega$ and $\omega$ is a positive real number (the \emph{frequency}). Note that then also $a_N(-i\omega)$ must equal the complex conjugate of $a_N(i\omega)$. In fact, using Lemma \ref{lem:taylor} in the previous Proposition yields that $a_N(0)=1$.  Thus it is sufficient to study $a_N(i\omega)$ only for $\omega> 0$.

\vskip .2in
\begin{prop}
Suppose $f$, $g$, $\omega>0$ and $\rho\in (0,1/2)\cup(1/2,1)$ are all fixed. Then, for $r\in(0,1)$ as in Lemma \ref{lem:mu+bigger}, as $N$ large tends to infinity:
\bsenn
a_N(i\omega)= \frac{1+\kappa}{\kappa}\;\mu_-^N\;\frac{\mu_+-\mu_-}{\mu_+-\mu_+^{-1}} \left(1+{\cal O}(r^N)\right)  \quad .
\esenn
\label{prop:rho-small-big}
\end{prop}

\noindent {\bf Proof:}
Use Proposition \ref{theo:a_n} and the fact that $\mu_-\mu_+=\kappa$ to rewrite
\bse
a_N(\nu)=\frac{1+\kappa}{\kappa}\;\mu_-^N\;\frac{\mu_+-\mu_-}{\mu_+-\mu_+^{-1}} \;\left(1- \frac{\mu_--\mu_-^{-1}}{\mu_+-\mu_+^{-1}} \left(\frac{\mu_-}{\mu_+}\right)^N\right)^{-1} \quad .
\label{eq:aN-to-mu}
\ese
Since
\bsenn
\frac{\mu_--\mu_-^{-1}}{\mu_+-\mu_+^{-1}}= \frac{-1}{\kappa}\; \frac{\mu_+^2-\kappa^2}{\mu_+^2-1} \quad ,
\esenn
it suffices to prove that if $\omega>0$, then $\mu_+(\omega)^2\neq 1$.
Now suppose that $\mu_+(\omega)^2= 1$, then the second remark after Proposition \ref{theo:a_n} implies that $\gamma(i\omega)=\pm1$ and so Lemma \ref{lem:gamma} implies that $\omega=0$. \QED

\noindent{\bf Remark:} It is important to realize that the even for large but \emph{fixed} $N$, the last factor could become large if we let $\mu_+$ approximate 1. The results in the appendix can be used to show that this can happen if and only if $\rho<\frac12$ and $\omega$ very close to zero. We make use of this fact in the proof (for $\rho<\frac12$) of the next and main result of this section.

\begin{theo} For all $\rho \in [0,1)\backslash \{\frac12\}$, $A_N$ grows exponentially in $N$. When $\rho=\frac12$, growth is linear in $N$.
\label{theo:stable2}
\end{theo}

\noindent {\bf Proof:}
The second statement has been proved in \cite{flocks5}. Fix $\rho\in(0,1/2)$ and let $\omega_+$ as in Equation \ref{eq:omega+}. Lemma \ref{lem:omega+} and the remark thereafter now imply that $|\mu_-(i\omega)|>1$ if and only if $\omega\in(0,\omega_+)$.
The result follows directly from the first part of Proposition \ref{prop:rho-small-big}. Finally fix $\rho\in (\frac12,1)$, or $\kappa\in (0,1)$.
First use Lemma \ref{lem:taylor} to see that if $\omega^2=\frac12 |f|(1-\kappa)^2\kappa^{N-1}$, then $\mu_+=1-\frac{(1-\kappa^2)}{2}\,\kappa^{N-1}+{\cal O}(\kappa^{3N/2})$ and $\mu_-=\kappa(1+\frac{(1-\kappa^2)}{2}\,\kappa^{N-1})+{\cal O}(\kappa^{3N/2})$.  Substitute this into the denominator of $a_N$ in Corollary \ref{cor:trailing}. The leading order cancels. The next term  is of order at least $\kappa^{3N/2}$. Thus $A_N$ is of order at least $\kappa^{-N/2}$.                  \QED

\vskip 1.in
\begin{centering}\section{Appendix: Technical Results} \label{chap:tresults} \end{centering}\
\setcounter{figure}{0} \setcounter{equation}{0}

For completeness we collect a number of straightforward results that are necessary for development of the theory, but would clutter the exposition in the main text.
Various relevant quantities are evaluated for $\nu=i\omega$ where $\omega$ is real and non-negative. As observed in the main text, we use $\omega\geq 0$ without loss of generality. The conventions mentioned in the Introduction also hold.

\vskip .2in
\begin{lem} i): For $\rho \in (0,\frac12)$ and $\omega \geq 0$ small:
\bsenn
\mu_+ = \frac{1-\rho}{\rho}\left(1 + \frac{\omega^2}{(2\rho-1)|f|} - i\; \frac{|g| \;\;\omega^3}{(2\rho-1)f^2} \right) + {\cal O}(\omega^4)  \quad \logand \quad
\mu_- =  1 - \frac{\omega^2}{(2\rho-1)|f|} + i\; \frac{|g| \;\;\omega^3}{(2\rho-1)f^2} + {\cal O}(\omega^4)  \quad .
\esenn
ii): For $\rho \in (\frac12,1)$ and $\omega \geq 0$ small:
\bsenn
\mu_+ = 1 - \frac{\omega^2}{(2\rho-1)|f|} + i\; \frac{|g| \;\;\omega^3}{(2\rho-1)f^2} + {\cal O}(\omega^4)  \quad \logand \quad
\mu_- =  \frac{1-\rho}{\rho}\left(1 + \frac{\omega^2}{(2\rho-1)|f|} - i\; \frac{|g| \;\;\omega^3}{(2\rho-1)f^2} \right) + {\cal O}(\omega^4)  \quad .
\esenn
\label{lem:taylor}
\end{lem}

\noindent {\bf Proof:} By sheer calculation. (See \cite{flocks4} for some of the computational details.)             \QED

\noindent {\bf Remark:} This expansion diverges for $\rho=1/2$; in that case we have (\cite{flocks4}):
 \bsenn
\mu_\pm = 1-\frac{\omega^2}{|f|}\pm
\frac{\omega^2|g|}{\sqrt{2}|f|^{3/2}}+{\cal O}(\omega^4) +i\left( \pm
\frac{\sqrt{2}\omega}{|f|^{1/2}} \pm {\cal O}(\omega^3) \right)
\quad .
 \esenn

\begin{lem} $\dpy\gamma(i\omega)=1-\frac{\omega^2 |f|}{f^2+\omega^2g^2} + i\,\frac{\omega^3|g|}{f^2+\omega^2g^2}\quad$.
\label{lem:gamma}
\end{lem}

\begin{figure}[ptbh]
\centering
\includegraphics[height=2.3in]{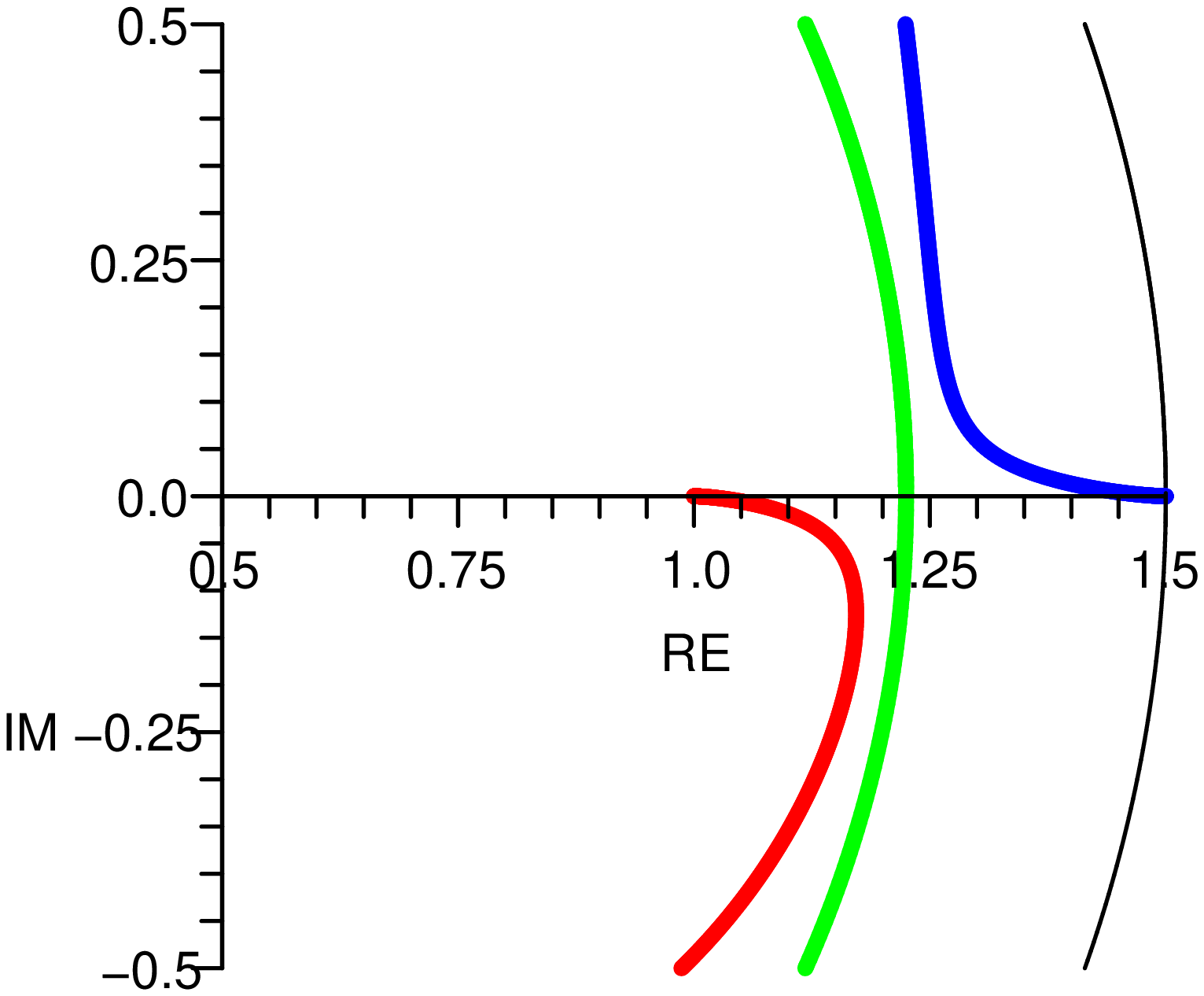}
\includegraphics[height=2.3in]{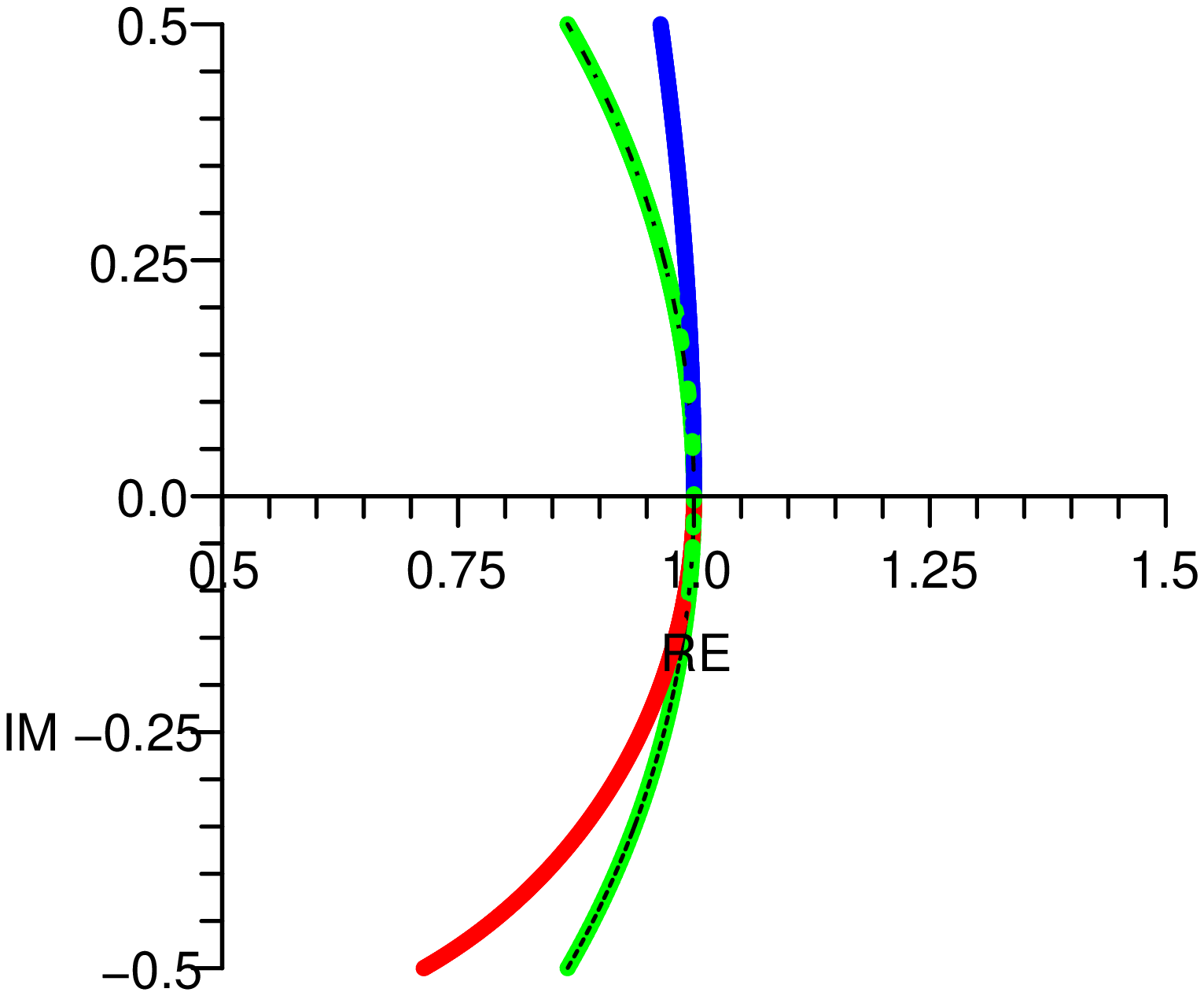}
\includegraphics[height=2.3in]{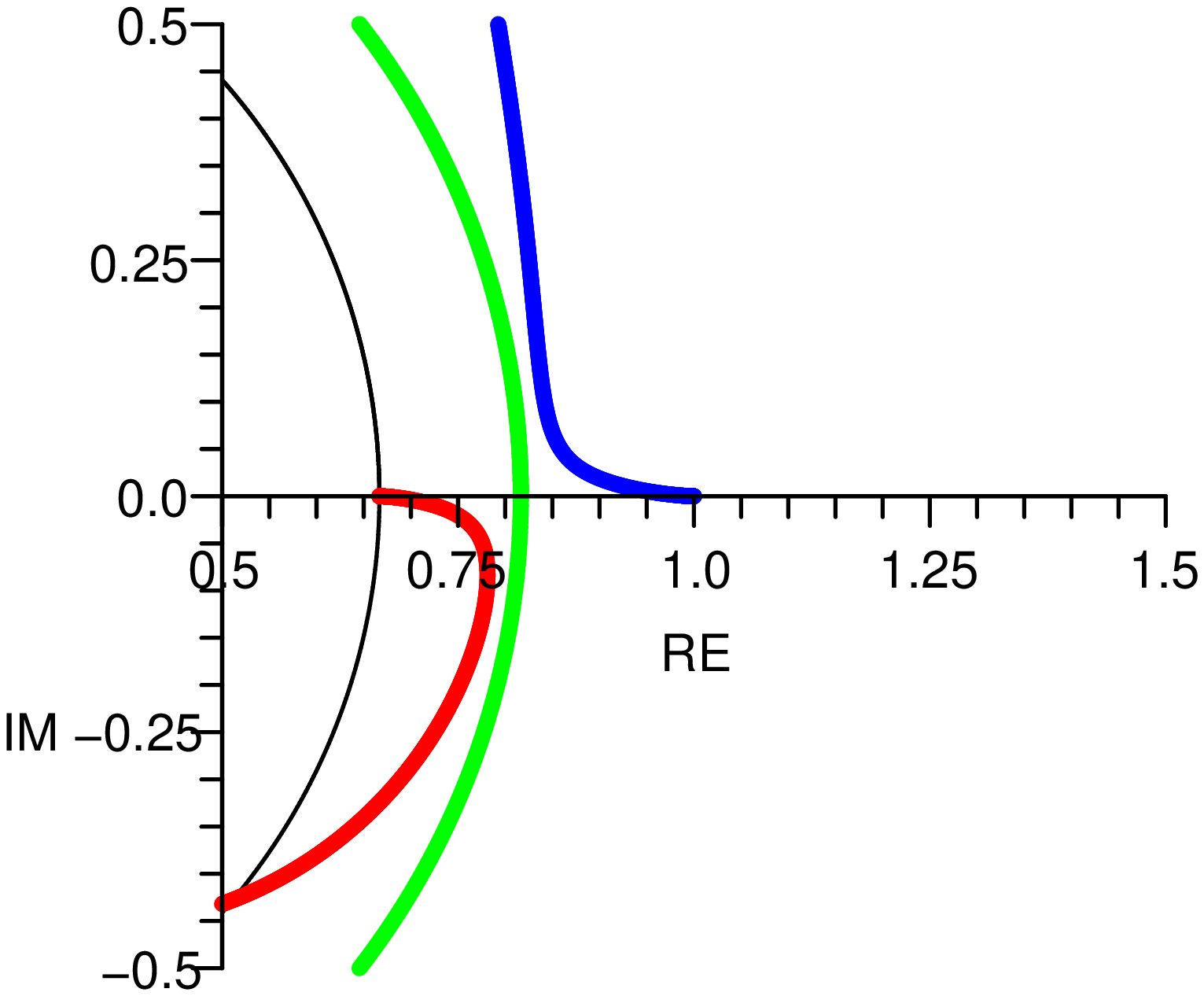}
\caption{\emph{ The eigenvalues $\mu_+(i\omega)$ (blue) and $\mu_-(i\omega)$ (red) of $C$ when $f=g=-1$ for $\omega$ positive.
From left to right: $\rho=0.4$, $0.5$, and $0.6$. In addition the circles with radii $\sqrt{\kappa}$ and $\kappa$ are drawn in green and black, resp., where $\kappa\equiv \frac{1-\rho}{\rho}$.}}
\label{fig:rho}
\end{figure}

\begin{lem} For each $\rho \in (0,1)\backslash \{\frac12\}$, the number $r\equiv \sup_\omega \frac{|\mu_-(i\omega)|}{|\mu_+(i\omega)|}$ exists and is contained in $(0,1)$. (See Figure \ref{fig:rho}.)
\label{lem:mu+bigger}
\end{lem}

\noindent {\bf Proof:} From Lemma \ref{lem:gamma}, $\gamma(i\omega)\approx -\frac{i\omega}{g}$ when $\omega$ is large. Substitute this into the expression for $\mu_\pm$ in Theorem \ref{theo:a_n} to see that for large $\omega$, in fact $\frac{|\mu_-(i\omega)|}{|\mu_+(i\omega)|}$ becomes very small.
When $\omega=0$, Lemma \ref{lem:taylor} implies that $\frac{|\mu_-(i\omega)|}{|\mu_+(i\omega)|}= \min\{\kappa,\kappa^{-1}\}$.

It is now sufficient to prove that for $\omega\in\R^+$ the absolute values $|\mu_\pm|$ are never equal. So suppose there are $\omega_0$ and  $\theta\in\R$ so that $\mu_+(i\omega_0)-\mu_-(i\omega_0)e^{i\theta}=0$. The first item in Theorem \ref{theo:a_n} gives:
\bsenn
\gamma (1-e^{i\theta}) = -\sqrt{\gamma^2-4\rho(1-\rho)}\,(1+e^{i\theta})\quad \quad .
\esenn
Dividing this by $1+e^{i\theta}$, squaring the equation, and noting that $\frac{(1-e^{i\theta})^2}{(1+e^{i\theta})^2}=-(\tan(\frac\theta2))^2$, we see that
\bsenn
\gamma^2 \left(1+\left(\tan \frac{\theta}{2}\right)^2\right)=4\rho(1-\rho) \quad .
\esenn
This implies that $\gamma^2$ is a positive real and therefore $\gamma$ is real for some $\omega\neq 0$, which is impossible by Lemma \ref{lem:gamma}.
\QED

\begin{lem} For each $\rho \in (0,1/2)$, there is a unique $\omega_+>0$ such that
\bsenn
\omega\in\left(0,\omega_+\right)\quad \Longrightarrow \quad |\mu_-(i\omega)|>1  \quad  \logand \quad \omega>\omega_+\quad \Longrightarrow \quad|\mu_-(i\omega)|<1 \quad .
\esenn
\label{lem:omega+}
\end{lem}

\noindent {\bf Proof:} We know that $\mu_-(0)=1$ and (from the proof of the previous Lemma) for large $\omega$: $|\mu_-(\omega)|$ is small. It is sufficient to prove that $\omega_+$ is the unique solution in $(0,\infty)$ of $|\mu_-(i\omega)|=1$ and that it is simple.

Consider the characteristic equation
$\rho\mu^2-\gamma \mu + (1-\rho)=0$  and suppose that there is a root $\mu=e^{i\theta}$. Then $\gamma=\rho e^{i\theta}+(1-\rho)e^{-i\theta} =\cos(\theta)+i(2\rho-1)\sin(\theta)$. Equate this to the expression given in Lemma \ref{lem:gamma} and use the fact that $\cos^2(\theta)+\sin^2(\theta)=1$ to obtain:
\bsenn
\left(1-\dfrac{\omega^2 |f|} {f^2+\omega^2g^2}\right)^2+\dfrac{1}{(2\rho-1)^2}\left(\dfrac{\omega^3|g|} {f^2+\omega^2g^2}\right)^2=1
\esenn
This equation factors as follows:
\bsenn
\omega^2\left(\dfrac{g^2}{(2\rho-1)^2}\,\omega^4+(f^2-2|f|g^2)\omega^2-2|f|^3 \right)=0
\esenn
The second factor gives exactly one simple positive root for $\omega^2$, yielding a unique simple positive root $\omega=\omega_+$.         \QED

\noindent {\bf Remark:} In fact,
\bse
\omega_+^2= (1-2\rho)|f|\left( \left(1-\frac{|f|}{2g^2}\right)(1-2\rho) + \sqrt{\left(1-\frac{|f|}{2g^2}\right)^2(1-2\rho)^2 +\,\frac{2|f|}{g^2}} \right) \quad .
\label{eq:omega+}
\ese

\vspace{\fill}

\end{document}